# AlCrO protected textured stainless steel surface for high temperature solar selective absorber applications

Vasiliy Pelenovich [*, 1)], Xiaomei Zeng [2)], Yan Liu [2)], Xiangyu Zhang [2)], Huidong Liu [2)], Alexander Pogrebnjak [3)], Ramil' Vildanov [4)], Bing Yang [2,5)], Sheng Liu [1)]

[1] Institute of Technological Sciences of Wuhan University, Wuhan, 430072, China

[2] School of Power & Mechanical Engineering, Wuhan University, Wuhan, 430072, China

[3] Sumy State University, Sumy, 40007, Ukraine

[4] Faculty of Physics, National University of Uzbekistan, Tashkent, 100174, Uzbekistan

[5] Shenzhen Research Institute of Wuhan University, Shenzhen, 518057, China

E-mail: pelenovich@mail.ru

* Corresponding author

## Abstract

The diffusion of substrate material into absorbing layer and oxidation of metal substrate or cermet metal nanoparticles at high temperatures are known as inevitable problems of the solar selective absorbers. In this study, we consider the use of textured stainless steel (SS) surface coated with a protective AlCr oxide layer as a high temperature solar selective absorber. The textured SS surface was prepared by ion etching techniques and AlCr oxide protective layer was deposited by RF magnetron sputtering. The absorptivity and emissivity of the as-prepared absorbers were 0.86-0.92 and 0.151-0.168, respectively. In order to evaluate the thermal stability, the absorbers were annealed at 600-800 °C for different time in ambient atmosphere. Absorbers demonstrated a red shift of the onset of the reflectivity at all annealing temperatures. The high activation energy of 315 kJ/mol was calculated. The service lifetime of the absorbers at 500 °C was estimated to be about 100 years and at 700 and 800 °C the absorbers were stable about 50 and 1 hours, respectively. A detailed examination of the annealed absorber surface revealed growth of surface $Mn_3O_4$ nanocrystals, which resulted in observed change of the reflectance spectra, while the textured surface morphology had no significant change. The results show that the protective textured surface has much higher thermal stability in air than iron based cermet absorbers.

# 1. Introduction

The use of solar selective absorber (SSA) is the most direct way to convert solar radiation into heat energy [1-6]. From the perspective of thermodynamics, in order to improve the efficiency of solar power generation system, it is necessary to use concentrated solar radiation [7]. The radiation concentration on the surface of the absorber will inevitably lead to the increase of its working temperature. Nowadays, for concentrating solar power generation systems, based on parabolic trough, temperatures of higher than 500 °C are required. Due to the strict requirements on the long-term thermal stability of absorbing materials, the design of selectively absorbing materials that can work at such temperature is a very difficult problem [2-6]. One of the failure mechanism at high temperature operation in air atmosphere is the oxidation of absorption layer and surface of metal infrared (IR) mirror, which leads to change of the optical properties of the solar absorber [8]. Another mechanism is the diffusion of atoms between the absorber sublayers (antireflection (AR) layer, absorbing layer, and IR mirror) at high temperature, which results to the loss of the SSA structure [9].

Both processes can be described by the diffusion equation:

$$\frac{\partial c}{\partial t} = D\Delta c \qquad (1)$$

where $c$ is the atomic concentration of metal or oxygen, $t$ is the time, $D$ is the diffusion coefficient, and $\Delta$ is the Laplacian [10]. The characteristic of any solution of the diffusion equation is the relaxation time $\tau$. It describes the duration of the diffusion process or, in this case, the duration of complete oxidation or structural loss of the absorber. The relaxation time can be written as:

$$\tau = \frac{l^2}{D} \qquad (2)$$

where $l$ is the size of the region where the atomic concentration reaches equilibrium [10]. It can be seen from this formula that the larger the diffusion region the longer the diffusion process, therefore, the longer the absorber can withstand high temperature.

At present, the most widespread SSAs are cermet absorbers, which are composed of 5-80 nm absorbing metal particles embedded in the dielectric matrix. The others are multilayer (dielectric-metal-dielectric) interference stacks consisting of 5-20 nm absorbing metal layer between two transparent dielectric layers. The textured metal surface can also be used as high temperature absorber if the size of the surface features $d$ is comparable with the wavelength of visible light (about 500 nm). In this case the wavefront discrimination effect takes place. If the wavelength of the incident radiation $\lambda$ is shorter then the size $d$, it is effectively multiply scattered from the surface resulting in enhancement of the absorptivity. If $\lambda \gg d$ the textured surface should be considered as a smooth reflective surface [11].

The failure of cermet absorber or interference stack at high temperature can be regarded as the diffusion/oxidation of metal particles/layers in oxide matrix. Whereas, the textured surface will lose its selective absorption, if the surface features change their morphology due to diffusion/oxidation process, that is, the average size of the feature is not longer equal to 500 nm. The comparison of the characteristic sizes $l$ of the absorbing elements discussed above shows that if the diffusion coefficients $D$ of metal (or oxygen) of all three absorber structures are the same, then according to Eq.2, the textured surface has the maximum relaxation time, i.e. it can be used for a longer time. Therefore, it can be concluded that the textured surface (protected by a suitable dielectric layer) design can be more suitable for high temperature applications.

In this study we do not consider the effect of material dependent diffusion coefficient $D$ on the service lifetime of the absorber, see Eq.2, therefore, further, we only focus on previous studies on stainless steel (SS) or iron-based cermet absorbers with $Al_2O_3$ or AlN dielectric matrix, where the SS or Fe are in the form of nanoparticles. SS as a construction material is very widespread and relatively cheap, therefore, is very attractive to be used for solar absorbers.

Using RF magnetron sputtering Sella et al. first successfully synthesized thin Fe-$Al_2O_3$ nanocomposite film overcoated with $Al_2O_3$ AR coating [12]. Fe nanoparticles mean size was found in the range of 40-120 nm. The solar absorptivity was 0.95 and the emissivity was 0.06 (Mo IR mirror). Thermal stability was studied in the Ar/$H_2$ atmosphere. The films deposited on SS and Nicral substrates showed good stability at 500 and 600 °C for 36 h, respectively. SS-AlN cermet selective surface with the double cermet layer structure was studied by Zhang [13]. A solar absorptivity of 0.93-0.96 and emissivity of 0.03-0.04 (Al IR mirror) have been achieved. However, thermal stability was studied only in vacuum at 500 °C for 1 h. In another work study of high temperature performance in air of Cu/SS-AlN/SiAlO$_x$ absorber has shown thermal stability of the material at 550 °C for about 40 h [14]. There are a few works on preparation and study of thermal stability of unprotected textured SS surface. Harding and Lake using sputter etching prepared textured surface on SS [15]. Its absorptivity reached value of 0.93 but at the expense of a rather high emissivity of 0.23. The selective surface was stable at 400 and 500 °C in air and vacuum, respectively. Similar thermal stability in air but worse optical performance were obtained for Fe textured surface prepared by catalyst techniques [16]. Through electrochemical treatment a self assembled 2D nanostructure was constructed on the Fe–Cr–Ni alloy surface and good selective light absorption was achieved [17]. A SSA with foamed nanostructure can also be grown by facile hydrothermal method on SS [18]. The foamed absorber shows good absorptivity and emissivity of 0.92 and 0.12, respectively, and thermal stability in air at 450 °C for a few tens hours. The analysis of literature data shows that there is a lack of information on high temperature stability in air of SS and iron based cermet absorbers, moreover, there is very little mention of using protective layers to improve high temperature performance of bare etched SS surface. It can be concluded that SS cermets and etched SS surface cannot be thermally stable in air at temperatures higher than 600 and 500 °C, respectively.

In this work, in order to verify the superiority of using protected textured surface for high temperature SSAs, we prepare textured SS surface by ion etching method and protect it by AlCr oxide layer deposited by RF magnetron sputtering, which acts as AR layer at the same time. We study the thermal stability of the prepared absorbers in air at temperatures up to 800 °C as well as mechanisms of their degradation. The protective AlCr oxide has been selected because of its good performance at temperatures up to 600 °C, which is proved in our previous papers [9,19].

## 2. Experimental details

Textured surface on 50 um thick SS foil was prepared by ion etching in 0.6 Pa Ar atmosphere using arc-enhanced glow discharge process, with negative bias voltage of 200 V at 80% duty cycle for 30-60 min. Next, an AlCr oxide protective/AR layer was deposited by reactive RF magnetron sputtering at 800 W power to coat the textured SS surface. The composition of 6-inch AlCr target was 70:30 at.%. The coatings were prepared in $O_2$/Ar atmosphere with flux ratio of 1:1 at a pressure of 1.5 Pa for 15-20 min and at a target-sample distance of 75 mm.

The crystal structure of the absorbers was characterized by X-ray diffraction (XRD) technique using D8 Advance diffractometer with a Cu Kα ($\lambda$ = 1.54 Å) source. The surface morphology was examined using a TESCAN MIRA3 scanning electron microscope (SEM) operated at 20 kV. The element composition and surface mapping were studied with Oxford Instruments X-MAX$^N$ detector for energy dispersive X-ray spectroscopy (EDS) operated at 20 kV. The surface morphology of the samples also was analyzed by atomic force microscopy (AFM, Shimadzu SPM-9500J3) operated in tapping mode with a measuring area of 5×5 um$^2$. Details of optical measurements at room temperature can be found in our previous study [20]. The thermal stability test was carried out in the resistance furnace in air at 600-800 °C for 5 min to 2000 hours. During the annealing process, the samples were taken out periodically for optical measurements.

To estimate the durability of absorbers a performance criterion (PC) was used, PC = -Δα + 0.5Δε$_{298K}$, where Δα and Δε$_{298K}$ are the changes of the solar absorptivity and thermal emissivity after thermal stability test, respectively [21]. The lifetime estimation of absorber at a certain temperature is based on the Arrhenius relation [22]:

$$a_n = \frac{t_r}{t_n} = \exp\left[\frac{E_a}{R}\left(\frac{1}{T_r} - \frac{1}{T_n}\right)\right] \tag{3}$$

where $a_n$ is an acceleration factor, $t_r$ and $t_n$ are time intervals under the reference temperature load and any other necessary temperature load with the same degradation level, respectively, $E_a$ is the activation energy, $R = 8.31$ J/K·mol is the gas constant, and $T_r$ and $T_n$ are the reference and any other necessary temperatures, respectively. If the $E_a$ value is unknown, it can be obtained from the same equation:

$$E_a = R \ln\left(\frac{t_r}{t_n}\right) \frac{T_n T_r}{T_n - T_r} \tag{4}.$$

It is seen that the calculation of $E_a$ requires at least two accelerated aging tests at different temperatures.

## 3. Results and discussion

**Figs.1a** and **1b** show SEM images of bare etched and coated with AlCr oxide protective layer SS surfaces, respectively. Well developed particulate morphology can be observed on the surface with a characteristic size of features about 300-500 nm, which is suitable for solar radiation absorption. **Fig.1c** demonstrates a typical reflectance spectrum of the bare etched substrate surface (black line) with the absorptivity in the range of 0.69-0.81 and the thermal emissivity at room temperature in the range of 0.136-0.164 depending on the etching time. The emissivity of the etched surface is higher than that of initial SS surface, $\varepsilon_{298K} = 0.122$, its corresponding reflectance spectrum is also shown in **Fig.1c** (blue line). The deposition of AlCr oxide AR layer demonstrates improving of the optical performance, as shown in **Fig.1c** (red line). In order to adjust the reflectance minimum at 500-700 nm to obtain the best absorptivity, the thickness of protective oxide layer was selected in the range of 70-100 nm. The absorptivity and emissivity of the absorbers with AlCr oxide protective layer are

0.86-0.92 and 0.151-0.168, respectively.

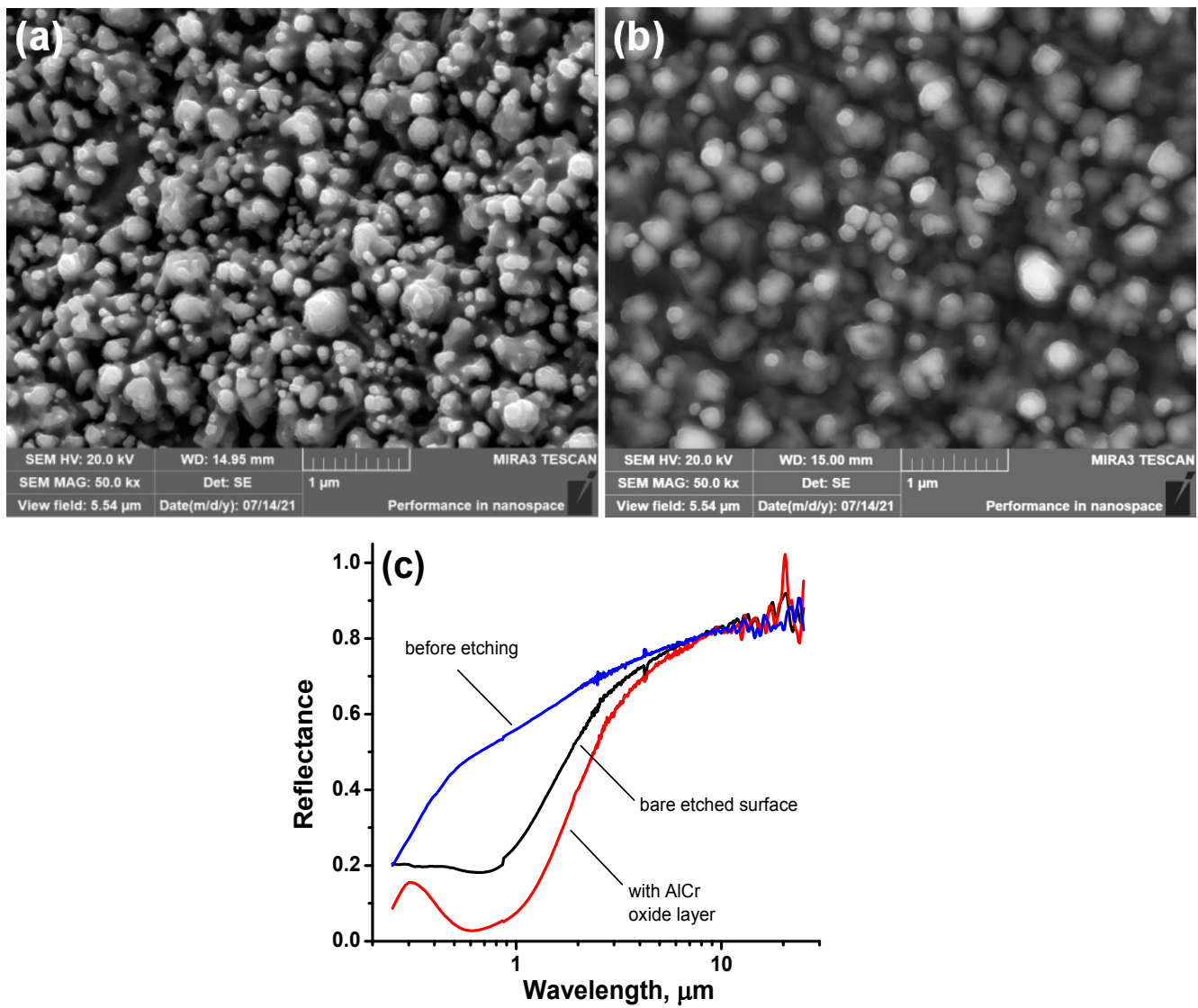

**Fig.1**. SEM images of bare etched SS substrate (a) and coated with AlCr oxide protective layer (b). Reflectance spectra of initial (blue line), bare etched (black line), and coated with AlCr oxide SS surfaces (red line) (c).

Next, we have carried out thermal stability tests in the air in the temperature range of 600-800 °C. The reflectance spectra of two absorbers during the aging tests at 700 and 800 °C are shown in **Fig.2a** and **b**. The solar absorptivity and thermal emissivity are shown in the insets of **Fig.2a** and **b**. At all temperatures, the absorbers exhibit similar change of reflectance spectra during annealing. In the ultraviolet region, the reflectance decreases significantly, whereas in the visible region a slight increase is observed. The most obvious change of the spectra is a red shift of the reflectivity onset in the range of 1-4 um. This change of the reflectance results in increase in both solar absorptivity and emissivity, as seen in insets. The PC values of all samples are negative, about 3-4%, indicating that the spectral selectivity is improved.

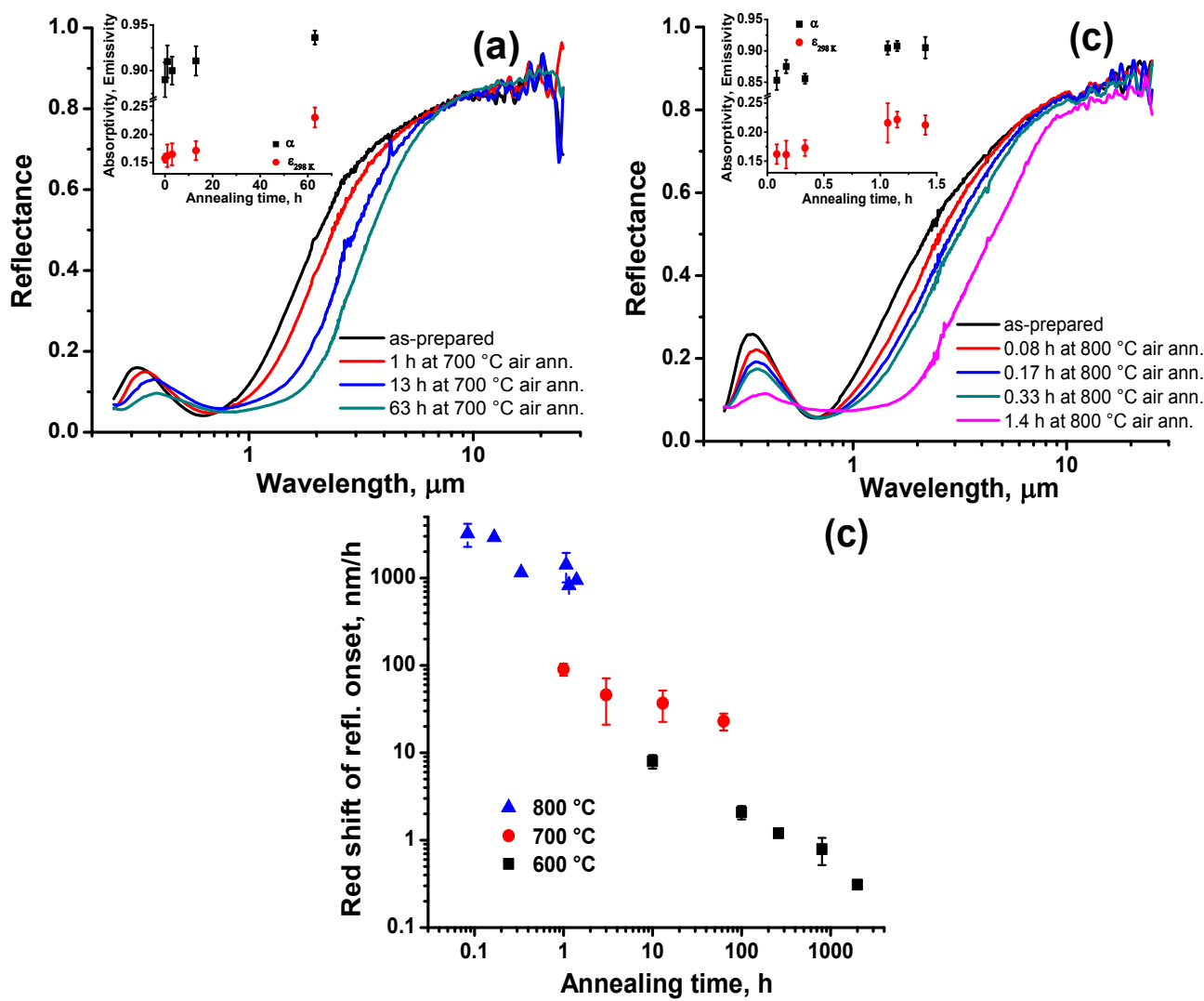

**Fig.2**. The reflectance spectra after different annealing times in air at 700 °C (a) and 800 °C (b). Insets represent the absorptivity α and emissivity $\varepsilon_{298K}$ as functions of annealing time. The red shift rates of the reflectance onset during annealing at different temperatures as functions of annealing time (c).

The red shift rates of the reflectivity onset (in nm/h) as functions of annealing time at the reflectance value of 0.4 for different temperatures are shown in **Fig.2c**. We define the red shift rate as $k = \Delta\lambda/t$, where $\Delta\lambda$ is a red shift in nm and $t$ is the annealing time interval. At all temperatures in the range of 600-800 °C the rate decreases with annealing time. The lifetime of absorbers at 400 and 500 °C can be estimated by using Eq.3. But the activation energy $E_a$ needs to be calculated first. Since the red shift rate $k$ is inversely proportional to the annealing time interval $t$, the acceleration factor can be rewritten as:

$$a_n = \frac{t_r}{t_n} = \frac{k_n}{k_r} \quad (5)$$

where $k_n$ and $k_r$ are the shift rates at two different temperatures and the same degradation level, that is, at the same red shift $\Delta\lambda$. Now, using results of two tests at 700 and 800 °C we can find $E_a$ value of 326±27 kJ/mol. Similar calculation for 600 and 700 °C gives $E_a$ = 303±44 kJ/mol. The obtained values are very close to each other, which proves the applicability of Arrhenius law in this temperature range. It should be noticed that the calculated $E_a$ values are greater than those in the range of 60-270 kJ/mol of nanocomposite absorbers, interference stacks and selective paint coatings [23-25]. Therefore, it can be concluded that AlCr oxide protective coating has excellent diffusion barrier properties. The surface life of the absorbers can now be estimated at 400 and 500 °C. Taking surface life of 1 h at 800 °C and average $E_a$ of 315 kJ/mol the surface life of about $1.5 \cdot 10^5$ and 100 years at 400 and 500 °C, can be obtained, respectively, which is much greater than the minimum service lifetime of 25 years.

It can be concluded, that the textured SS surface protected with AlCr oxide layer, in terms of PC value, can be stable at 700 and 800 °C for more than 50 and 1 h, respectively. The obtained thermal stability is considerably higher than those of bare etched SS surface and Fe or SS based cermets, where degradation of the optical properties was found after annealing in air at 500 and 600 °C for tens hours, respectively.

XRD patterns of the absorber and unprotected etched SS surface before and after annealing at 800 °C for 1.4 h are shown in **Fig.3a**. The XRD pattern of the as prepared absorber has not revealed any peaks of protective layer indicating its amorphous or fine nanocrystalline structure. Both annealed absorber and unprotected SS surface demonstrate new phases with crystal structure of corundum (α-$Fe_2O_3$ or $Cr_2O_3$) and traces of phase with spinel structure ($Fe_3O_4$ or $Mn_3O_4$). One also can notice the stronger signal of oxides for unprotected SS surface, therefore, it can prove effectiveness of the AlCr oxide protective layer as the oxidation/diffusion barrier. SEM image of the absorber surface after annealing at 800 °C in air for 1.4 h also confirms formation of surface nanocrystals with a typical size of 500 nm, as shown in bottom-right corner of **Fig.3b**.

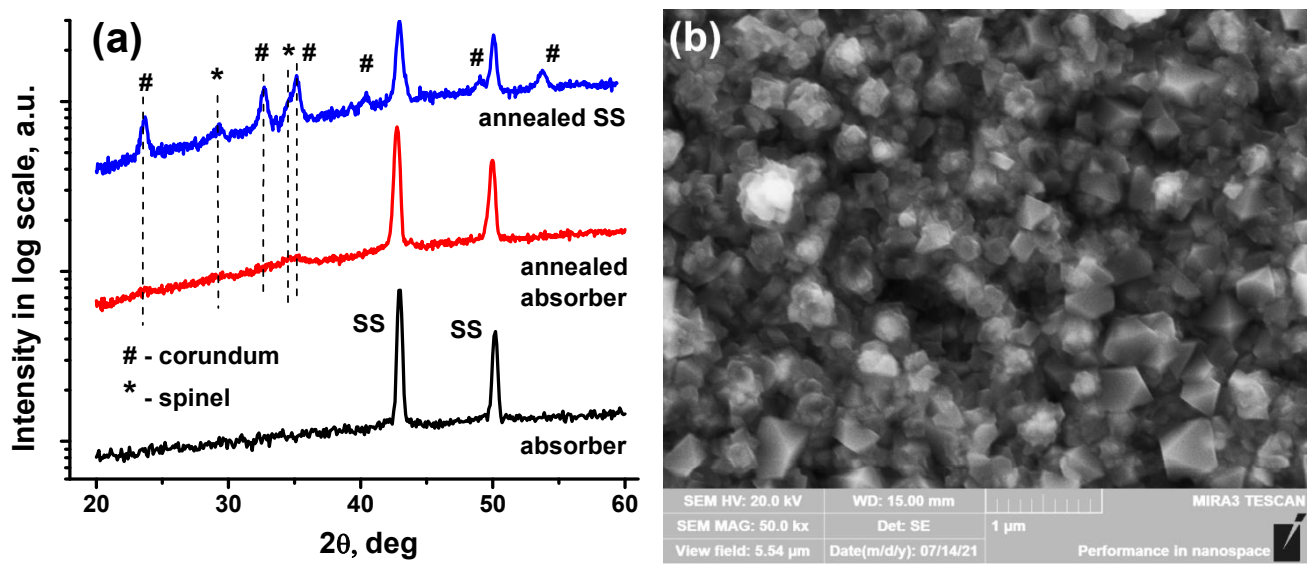

**Fig.3**. XRD patterns of the bare etched SS substrate and absorber before and after annealing at 800 °C for 1.4 h (a). SEM image of the absorber surface after annealing at 800 °C in air for 1.4 h (b).

If to compare absorber surface before annealing, **Fig.1b**, and after annealing at 800 °C for 1.4 h, **Fig.3b**, one can notice that the particulate texture of the surface still remains after annealing, but the surface of the particles is no longer smooth, however, the size of the particles does not change significantly. The change of morphology can be explained by the crystallization of initially amorphous AlCr oxide layer as well as oxidation of SS under the AlCr coating and formation of hematite crystals. The effect of annealing on particle morphology in more details can be observed in AFM images, shown in **Fig.4a** and **b**. It can be concluded that the textured surface itself is quite robust at 800 °C for about one hour and can still be used as an absorbing element of the solar absorber. After annealing an increase of RMS roughness from 110 to 135 nm is observed. Surface nanocrystals also can be seen on the right side of **Fig.4b**.

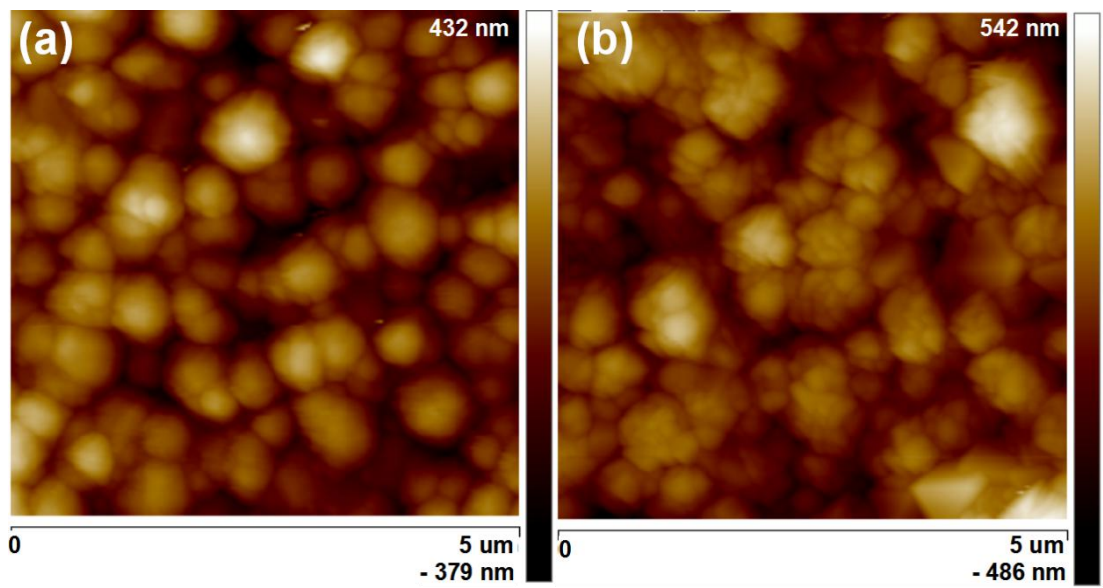

**Fig.4**. AFM images of absorber surfaces before (a) and after annealing at 800 °C for 1.4 h (b).

To further clarify the mechanism of high temperature degradation we have analyzed element composition of the absorbers before and after annealing at 800 °C in air for 1.4 h using EDS technique, see **Table 1**. Composition of the SS substrate before and after ion etching is also shown. It can be seen that etching does not significantly change the composition of the substrate. Coated with AlCr oxide layer SS sample still has signal from substrate (Si, Mn, Fe, Ni) due to insufficient thickness of the AlCr oxide layer. After annealing considerable increase and decrease of oxygen and iron contents are observed, respectively. Therefore, it can be concluded that SS substrate is oxidized under the AlCr oxide layer, which is in agreement with XRD data. It also can be noticed an increase of Si and especially Mn contents. Since only the SS substrate contains Mn and Si it should be concluded, that during annealing, both elements diffuse to the surface and oxidize on it.

**Table 1**. Element composition of not etched, bare etched, and coated with AlCr oxide protective layer SS surface before and after annealing at 800 °C in air for 1.4 h.

| Sample | O | Al | Si | Cr | Mn | Fe | Ni |
|---|---|---|---|---|---|---|---|
| SS substrate | | | 1.1 | 20.3 | 1.3 | 69.8 | 7.5 |
| bare etched SS | | | 0.8 | 19.6 | 1.4 | 71.6 | 6.6 |
| absorber | 20.4 | 8 | 0.5 | 15.9 | 0.9 | 49.8 | 4.5 |
| annealed absorber | 47.6 | 5.7 | 1 | 13.4 | 3.7 | 26.1 | 2.5 |

To study the composition of the grown surface nanocrystals after annealing at 800 °C observed in SEM and AFM images, we have conducted EDS element mapping measurement of a selected area shown in **Fig.5a**. In order to increase the measurement sensitivity, this sample was annealed for 30 h, in order to grow bigger surface crystals with a typical size of more than 1 um. **Figs.5b-d** show signal from O, Mn, and Fe elements. Comparison of the SEM and EDS images reveals that the nanocrystals consist of manganese oxide, according to XRD data, possibly $Mn_3O_4$. Manganese oxide nanocrystals can also be seen in the background of the iron signal, see **Fig.5d**.

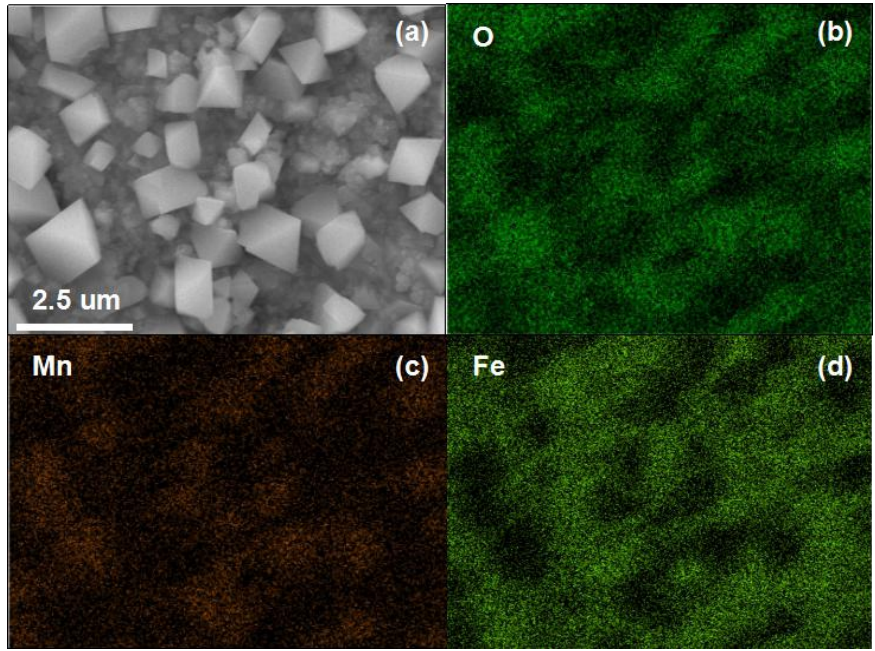

**Fig.5**. SEM image of the absorber surface after annealing at 800 °C in air for 30 h selected for EDS mapping (a). The element mapping images of O, Mn, and Fe (b-d).

**Fig.6a** and **b** show absorber surfaces after annealing at 800 °C in air for 1.4 and 30 h. An obvious growth of surface nanocrystals is observed. Corresponding reflectance spectra of the samples are shown in **Fig.6c** (black and red line), the reflectance spectra of the sample before annealing is also shown (blue line). Annealing at 800 °C in air for 30 h results in further red shift of the reflectivity onset and degradation of the optical performance with absorptivity of 0.898, thermal emissivity of 0.297, and PC value of more than 8 %. It can be concluded that there is a correlation between the continuous growth of the surface nanocrystals and gradient degradation of the absorber selectivity.

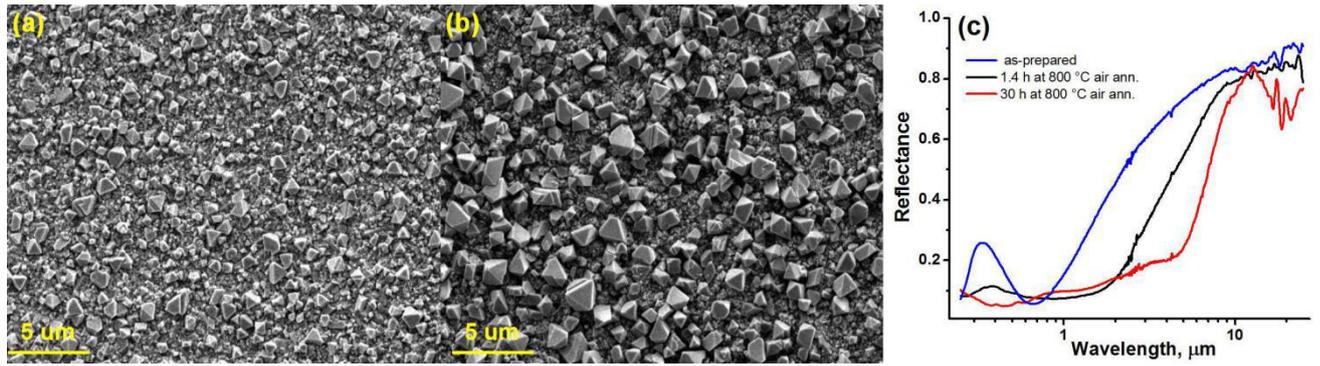

**Fig.6**. SEM images of the absorber surface after annealing at 800 °C in air for 1.4 (a) and 30 h (b) and corresponding reflectance spectra (c).

Now we can describe the degradation mechanism of the absorber during annealing, which is manifested in the red shift of reflectivity onset observed in the reflectance spectra. The interference maxima in the UV range demonstrate a small red shift, which means that the thickness of oxide layer on the surface does not change significantly during annealing. Whereas intensity of the UV maxima decreases, the interference minima are flattened, and the onset of reflectivity is shifted to longer wavelength increasing the emissivity, all these effects can be explained by the wavefront discrimination effect introduced by growing surface nanocrystals, which, therefore, act as unwanted textured surface. The contribution of the effect should be significant, since the area covered by the growing nanocrystals becomes comparable with the absorber surface area, see **Fig.6b**.

As shown by Rodríguez-Palomo et al. [26], it can be expected that the performance of the absorber can be improved, i.e. oxidation of the substrate material can be decreased, by using another high temperature substrate material instead of SS, for example, Mn free material.

In **Fig.2c** it can be seen that the rate of the onset shift decreases with time at all annealing temperatures, which indicates an increase of the absorber durability, possibly, due to strengthening of the oxidation barrier on the surface by formation of hematite layer under the AlCr oxide layer. Another explanation is related to the growth rate of $Mn_3O_4$ nanocrystals on the surface. If even linear diffusion and oxidation rate of the substrate material is considered, then the volume $V$ of nanocrystals is directly proportional to the annealing time $t$, and their linear size is $V^{1/3}$. The maximal wavelength $\lambda$, which can be effectively trapped by the textured surface composed of these nanocrystals, is proportional to their linear size, hence, during annealing $\lambda \sim V^{1/3} \sim t^{1/3}$, i.e. the time dependence is sublinear. Therefore, the rate of the red shift, which can also be defined as a derivative $d\lambda/dt$, decreases with time.

## 4. Conclusions

The analysis of diffusion equation reveals that the textured surface of solar absorber has higher durability at elevated temperatures than those of cermet and interference stack solar absorbers. In this study, the textured SS surface and top AlCr oxide protective layer have been prepared by ion etching in Ar atmosphere and reactive RF magnetron sputtering, respectively. The typical absorptivity and emissivity of the as-prepared absorbers are 0.86-0.92 and 0.151-0.168, respectively. The thermal stability in air of the textured absorbers in the temperature range of 600-800 °C has been studied. The absorbers are stable at 700 and 800 °C for about 50 and 1 h, respectively. The obtained thermal stability is considerably higher than those of previously studied bare etched SS surface and Fe or SS based cermets, where degradation of the optical properties was found after annealing in air at 500 and 600 °C for tens hours, respectively. During the annealing process all samples demonstrate shift of the reflectivity onset to the longer wavelength. This increases the absorptivity and emissivity up to 0.92-0.94 and 0.21-0.24, respectively. However, the PC values of all samples are negative, about 3-4%, indicating that the absorber spectral selectivity is improved. Using the red shift of the reflectivity onset as the degradation rate, the activation energy of 315 kJ/mol and the lifetime of the absorbers of ~100 years at 500 °C were calculated. A detailed examination of the annealed in air absorber surface revealed growth of surface $Mn_3O_4$ nanocrystals, resulting in the observed change of the reflectance spectra due to the wavefront discrimination effect. This factor can be the determinant of the absorber degradation mechanism. On the other hand, the texture morphology shows good thermal stability, therefore, it can be concluded that the protected texture surface is rather robust to high temperature annealing in air and can be suggested to use as a high temperature solar absorber instead of double layer cermet design. It can also be expected that the high temperature performance of the absorber can be significantly improved by using as the substrate another manganese free material instead of SS.

## Acknowledgments

This work was supported by the Science and Technology Program Project of Shenzhen (2019N032) and National Natural Science Foundation of China (U1832127).

## References

[1] C.G. Granqvist, Solar energy materials, Adv. Mater. 15 (2003) 1789–1803

[2] N. Selvakumar, H. C. Barshilia, Review of physical vapor deposited (PVD) spectrally selective coatings for mid- and high-temperature solar thermal applications, Sol. Energy Mater. Sol. Cells 98 (2012) 1–23

[3] C.E. Kennedy, Review of Mid- to High Temperature Solar Selective Absorber Materials, NREL/TP-520-31267, July 2002

[4] K. Xu, M. Du, L. Hao, J. Mi, Q. Yu, S. Li, A review of high-temperature selective absorbing coatings for solar thermal applications, J. Materiom. 6 (2020) 167-182

[5] K. Zhang, L. Hao, M. Du, J. Mi, J.-N. Wang, J.-P. Meng, A review on thermal stability and high temperature induced ageing mechanisms of solar absorber coatings, Renew. Sustain. Energy Rev. 67 (2017) 1282–1299

[6] M. Bello, S. Shanmugan, Achievements in mid and high-temperature selective absorber coatings by physical vapor deposition (PVD) for solar thermal Application-A review, J. Alloys Compd. 839 (2020) 155510

[7] C.H. Liebert, R.R. Hibbard, Performance of Spectrally Selective Collector Surfaces in a Solar-Driven Carnot Space-Power System, Solar Energy, 6(3) (1962) 84-88

[8] M. Du, X. Liu, L. Hao, X. Wang, J. Mi, L. Jiang, Q. Yu, Microstructure and thermal stability of Al/Ti$_{0.5}$Al$_{0.5}$N/Ti$_{0.25}$Al$_{0.75}$N/AlN solar selective coating, Sol. Energy Mater. Sol. Cells 111 (2013) 49–56

[9] H.D. Liu, Q. Wan, Y.R. Xu, C. Luo, Y.M. Chen, D.J. Fu, F. Ren, G. Luo, X.D. Cheng, X.J. Hu, B. Yang, Long-term thermal stability of CrAlO-based solar selective absorbing coating in elevated temperature air, Sol. Energy Mater. Sol. Cells 134 (2015) 261–267

[10] Fluid Mechanics. By L.D. Landau and E.M. Lifshitz. (Translated from the Russian by J. B. Sykes and W. H. Reid.) London: Pergamon Press, 1959, pp.194, 226.

[11] R.A. Buhrman, Physics of solar selective surfaces. In Advances in Solar Energy, Vol. 3, Boer K.W. (Ed.), ASES Plenum Press, New York, 1986, p.210

[12] C. Sella, A. Kaba, S. Berthier, J. Lafait, Low cost selective absorber based on a Fe-Al$_2$O$_3$ cermet film. SPIE Vol. 653 Optical Materials Technology for Energy Efficiency and Solar Energy Conversion V (1986) 208-214

[13] Q.-C. Zhang, Recent progress in high-temperature solar selective coatings, Sol. Energy Mater.

Sol. Cells 62 (2000) 63-74

[14] J.P. Wang, B. Yuan, J. Wang, Y. Luo, Activation Energy Measurement of Cu/SS-AlN/SiAlOx Solar Selective Absorbing Coating, Materials Science Forum 743-744 (2013) 870-877

[15] G.L. Harding, M.R. Lake, Sputter etched metal solar selective absorbing surfaces for high temperature thermal collectors, Sol. Energy Mater. 5 (1981) 445-464

[16] E. Erben, A. Muehlratzer, B.A. Tihanyi, B. Cornils, Development of high temperature solar selective absorbers utilizing rare earth, transitional, and group metals, Sol. Energy Mater. 9 (1983) 281-292

[17] D.W. Ding, W.M. Cai, Self-assembled nanostructured composites for solar absorber, Materials Letters 93 (2013) 269–271

[18] Wang X.L., Wu X., Yuan L., Zhou C., Wang Y., Huang K., Feng S., Solar selective absorbers with foamed nanostructure prepared by hydrothermal method on stainless steel, Sol. Energy Mater. Sol. Cells 146 (2016) 99–106

[19] H.D. Liu, T.R. Fu, M.H. Duan, Q. Wan, Structure and thermal stability of spectrally selective absorber based on AlCrON coating for solar-thermal conversion applications, Sol. Energy Mater. Sol. Cells 157 (2016) 108–116

[20] V. Pelenovich, H.D. Liu, X.M. Zeng, Y. Liu, K. Liu, B. Yang, Graded solar selective absorbers deposited by non-equilibrium RF magnetron sputtering, Sol. Energy Mater. Sol. Cells 230 (2021) 111188

[21] ISO 22975-3:2014. Solar Energy – Collector Components and Materials – Part 3: Absorber Surface Durability

[22] M. Kohl, K. Gindele, U. Frei, T. Hauselmann, Accelerated ageing test procedures for selective

absorber coatings including lifetime estimation and comparison with outdoor test results, Sol. Energy Mater. 19 (1989) 257-313

[23] H.C. Barshilia, N. Selvakumar, K.S. Rajam, A. Biswas, Optical properties and thermal stability of TiAlN/AlON tandem absorber prepared by reactive DC/RF magnetron sputtering, Sol. Energy Mater. Sol. Cells 92 (2008) 1425– 1433

[24] A. Dan, H.C. Barshilia, K. Chattopadhyay, B. Basu, Solar energy absorption mediated by surface plasma polaritons in spectrally selective dielectric-metal-dielectric coatings: A critical review, Renew. Sustain. Energy Rev. 79 (2017) 1050–1077

[25] R. Kunic, Mohor Mihelcic, B. Orel L. Slemenik Perse, B. Bizjak, J. Kovac, S. Brunold, Life expectancy prediction and application properties of novel polyurethane based thickness sensitive and thickness insensitive spectrally selective paint coatings for solar absorbers, Sol. Energy Mater. Sol. Cells 95 (2011) 2965–2975

[26] A. Rodríguez-Palomo, E. Céspedes, D. Hernández-Pinilla, et al., High-temperature air-stable solar selective coating based on $MoSi_2$–$Si_3N_4$ composite, Sol. Energy Mater. Sol. Cells 174 (2018) 50–55